\documentclass[seceq]{ptptex}

\usepackage{amsmath,amsfonts,latexsym,amssymb}
\usepackage{verbatim,amsthm,graphicx}
\usepackage{wrapft}
\usepackage{mathrsfs, amsmath}

\usepackage{bm}
\def\bm#1{{\boldsymbol{#1}}}

\usepackage{amsmath}
\usepackage{amssymb}
\usepackage{mathrsfs}

\def\ifempty#1{\def\tmpdata{#1}\ifx\tmpdata\empty }

\def\linebreak{\hfill\break}

\def\bra<#1|{\langle #1\rvert}
\def\ket|#1>{\lvert#1 \rangle}
\def\braket<#1|#2>{\langle #1|#2 \rangle}

\def\para{{\scriptscriptstyle /\!/}}

\def\otop#1{\hbox{$#1\kern-0.1em$\llap{\hbox{\raise1.7ex\hbox{$\scriptstyle\circ$}}}} }

\def\inpare#1{\left(#1\right)}
\def\bigpare(#1){\left(#1\right)}

\def\insbra#1{\left[ #1 \right]}
\def\inang#1{\left\langle {#1} \right\rangle}
\def\EXP#1{\inang{#1}}
\def\bigbra[#1]{\left[ #1 \right]}

\def\b{\bar }

\def\then{\Rightarrow\quad}

\def\therefore{\mbox{\setbox0=\hbox{X}\hbox{$\ldotp$}\raise0.7\ht0\hbox{$\ldotp$}\hbox{$\ldotp$}} \quad }
\def\because{\mbox{\setbox0=\hbox{X}\raise0.7\ht0\hbox{$\ldotp$}\hbox{$\ldotp$}\raise0.7\ht0\hbox{$\ldotp$}}\kern0pt }

\def\r#1{{\rm #1}}

\def\bm#1{\boldsymbol{#1}}

\def\upin{\hbox{\setbox0=\hbox{$\cup$} \vrule width 0.05 \wd0 height \ht0 depth 0pt \kern - 0.5\wd0 \box0 }}
\def\Frac(#1/#2){\left(\frac{#1}{#2}\right)}

\def\sdprod{\mathrel{{\setbox0=\hbox{$\displaystyle\times$}\lower0.3\wd0\hbox{$\stackrel{\box0}{\scriptstyle\sim}$}}}}

\def\tosigma#1,{%
    \ifx\tmpindex\relax \def\tmpindex{#1} \let\next=\tosigma
    \else \ifnum\tmpindex=0 1 \else \sigma_\tmpindex \fi
          \ifx#1\relax  \let\next=\relax
          \else \otimes \let\next=\tosigma \def\tmpindex{#1} \fi
    \fi \next}
\def\tspb(#1){\let\tmpindex=\relax\tosigma#1,\relax,}
\def\Order#1{{\rm O}\!\left(#1\right)}

\def\pd{\partial}

\def\HyperG(#1,#2;#3;#4){F\inpare{\textstyle #1,#2;#3;#4}}

\def\Eq#1{\begin{equation} #1 \end{equation}}

\def\Eqr#1{\begin{eqnarray} #1 \end{eqnarray}}

\def\Eqrsub#1{\begin{subequations}\Eqr{#1}\end{subequations}}
\def\Eqrsubl#1#2{\begin{subequations}
  \expandafter\ifx\csname Rlabel\endcsname \relax \label{#1}
  \else \Rlabel{#1} \fi \Eqr{#2}\end{subequations}}
\def\Bitm{\begin{itemize}}
\def\Eitm{\end{itemize}}
\def\Blist#1#2{\begin{list}{#1}{\parsep=0pt \itemsep=0pt%
  \listparindent=0pt #2}}
\def\Elist{\end{list}}
\long\def\ignore#1#2{\def\ignoreflag{#1}\long\def\tmptext{#2}
  \ifnum\ignoreflag>1 #2 \fi}

\begin{document}
%
\title{
The Gravitational Lensing Effect on the CMB Polarisation Anisotropy in the $\Lambda$-LTB Model
}
%
\author{%
  Hajime Goto\footnote{Email address: gotohaji@post.kek.jp}$^{(a)}$
  and
  Hideo Kodama\footnote{Email address: hideo.kodama@kek.jp}$^{(a),(b)}$
}
%
\inst{%
  $^{(a)}$Department of Particle and Nuclear Physics, Graduate University for Advanced Studies, Tsukuba 305-0801, Japan\\
  $^{(b)}$Theory Center, KEK, Tsukuba 305-0801, Japan}
%
\abst{

A local void modifies the sky distribution pattern of the cosmic microwave background (CMB) polarisation by gravitational lensing and produces B-modes from E-modes for an off-center observer. In order to see whether this effect can be used to observationally test the validity of the local void model, we calculate this lensing effect by solving the propagation of CMB polarisation along null geodesics close to the central light cone in the general Lema\^{i}tre-Tolman-Bondi (LTB) model perturbatively. In particular, we give general formulas for the correlations of $E$ and $B$ observed by an off-center observer and show that $E^m_\ell$ and $B^m_\ell$ are correlated for the same value of $\ell$, i.e., $\langle E^m_\ell B^{m'}_{\ell'}\rangle \propto \delta_{\ell,\ell'}$, while $\langle E^m_{\ell}E^{m'}_{\ell'}\rangle \propto \delta_{|\ell'-\ell|,1}$. This feature can be used to distinguish the gravitational lensing effect by a local void from those by normal shear field of galaxies.

}

\maketitle

\section{Introduction}
\label{sec:intro}

Type Ia supernova (SNIa) observations imply an acceleration of the cosmic expansion if general relativity is valid on cosmological scales and if the universe is homogeneous and isotropic on scales larger than 200Mpc. If we abandon one of these assumptions, however, other explanations become possible. One approach of such a nature is the so-called modified gravity that abandons general relativity. The other is the local void model, which was first proposed by Tomita \cite{2000ApJ...529...26T}\cite{2000ApJ...529...38T}, Goodwin et al. \cite{1999astro.ph..6187G}, and Celerier \cite{2000A&A...353...63C}, independently.  This model abandon the second assumption, which is often called the Cosmological Principle or the Copernican Principle, and assumes that we are around the center of a low density spherically symmetric void and that the spacetime is well described by the Lema\^{i}tre-Tolman-Bondi (LTB) model \cite{1933ASSB...53...51L}\cite{1934PNAS...20..169T}\cite{1947MNRAS.107..410B}. In this model, the cosmic expansion rate decreases outward on each constant time slice, which produces an apparent acceleration of the universe when observed along the past light cone. Although this model violates the Cosmological Principle and requires an accidental situation concerning our location in the universe, it does not require any dark energy or a modification of gravity theory. Further, as far as the redshift-luminosity distance relation obtained by the SNIa observations is concerned, this model can reproduce the observational results with any accuracy because it contains at least one arbitrary function of the radius (see, e.g., \citen{2008PThPh.120..937Y}). Actually, it has passed all observational tests so far. Therefore, it is of crucial importance to find observational tests that enable us to discriminate this void model from spatially homogeneous models employing dark energy or modified gravity theory, in order to establish the necessity of dark energy or a modification of gravity.

There have been proposed various tests for that purpose so far\cite{Tomita.K2009,Moss.A&Zibin&Scott2010}: CMB anisotropies on large angular scales, radial BAO\cite{2008JCAP...04..003G}, spectral distortions of CMB\cite{2008PhRvL.100s1302C}, the kinematic SZ effect\cite{2008JCAP...09..016G}, and estimates of the Hubble rate \cite{2010MNRAS.405.2231F}.
Among these, the oldest and simplest one is to observe the effect of the inhomogeneity on CMB temperature. This effect was first estimated by Alnes and  Amarzguioui\cite{AA2006} for a special class of void models\cite{AAG2006}.   
Recently, this analysis was extended to a wider class of models by Kodama, Saito and Ishibashi with the helps of analytic formulas for the dipole and quadrupole moments of the CMB temperature anisotropy for an off-center observer in a general spherically symmetric universe \cite{2010PThPh.124..163K}. Although this type of test provides a strong constraint on the allowed range of our distance from the void center, it is not so decisive because lower moments including the dipole and quadrupole are significantly affected by the cosmic variance. 

One possible way to circumvent this weakness is to extend the analysis of the off-center CMB anisotropies to polarisation. As is well known, in the spatially homogeneous cosmology, inhomogeneities of the matter (galaxy) distribution modify the sky pattern of CMB polarisation by gravitational lensing and produce B-modes from E-modes\cite{1996ApJ...463....1S}\cite{1998PhRvD..58b3003Z}\cite{2000PhRvD..62d3007H}\cite{2003PhRvD..67h3002O}\cite{2006PhR...429....1L}. In the local void model, the central observer detects no such effect because of the spherical symmetry, in spite of the strong inhomogeneity of the model. However,  an off-center observer can detect the gravitational shear field through the observation of B-modes.  Therefore, in this paper, we calculate the gravitational lensing effect on the CMB temperature and polarisation anisotropies for an off-center observer in the local void model and estimate the correlations among the temperature, the E-mode and the B-mode anisotropies.

The paper is organised as follows. First, in the next section, we review the basic matters on CMB polarisation that are relevant to the present paper. Then, in Section \ref{sec:gl}, we perturbatively solve the geodesic equations in the LTB model to find the change in the propagation direction due to the gravitational lensing effect for null rays close to the central past light cone. This determines the shift vector on the sky for an off-center observer that represents the difference between the observed direction of a light ray and the corresponding direction of the last scattering point in the fiducial spatially homogeneous model in a gauge in which the fiducial universe has the same spherical symmetry as the LTB background.  Next, in Section \ref{sec:polLTB}, we carefully examine how the polarisation evolves from the last scattering surface to the present time in the LTB spacetime with the help of the collisionless Boltzmann equation. In particular, we show that the temperature and polarisation anisotropy of CMB in the LTB model can be expressed by the same formula in terms of the shift vector as in the FLRW universe background. Putting these results together, in Section \ref{sec:result}, we give general formulas for the CMB temperature and polarisation anisotropy produced by the void inhomogeneity and for the correlations among the harmonic components of the temperature, the E-mode polarisation and the B-mode polarisation. Finally, Section \ref{sec:sum} is devoted to summary and discussions.

\section{CMB Polarisation Basics}
\label{sec:polbasic}

\subsection{How to Represent Polarisations}
\label{sec:rep}

First of all, we explain the standard method to represent the polarisation of radiations. Let us consider a quasi-monochromatic plane electromagnetic wave propagating toward an observer, and take an orthonormal $xy$-basis that is orthogonal to the wave propagation direction. Then, the electric field of the wave is represented as $\bm{E} = E_x \bm{e}_x + E_y \bm{e}_y$, with $E_x = a_x \sin (\omega t - \epsilon_x)$ and $E_y = a_y \sin (\omega t - \epsilon_y)$.

In this setup, one can define parameters that represent polarisation as follows: $I := \langle a_y^2 \rangle + \langle a_x^2 \rangle$, $Q := \langle a_y^2 \rangle - \langle a_x^2 \rangle$, $U := \langle 2 a_y a_x \cos (\epsilon_y - \epsilon_x) \rangle$, and $V := \langle 2 a_y a_x \sin (\epsilon_y - \epsilon_x) \rangle$.
These are called the Stokes parameters.
Physically, $I$ represents intensity (temperature), $Q$ and $U$ represent linear polarisation, and $V$ represents circular polarisation.
We ignore $V$ because circular polarisation is never generated by Thomson scattering in the early universe.

When the orthonormal basis is rotated in the wave plane, $Q$ and $U$ are linearly transformed. If we introduce the matrix $P$ defined by
\Eq{
P=\begin{pmatrix} Q & U \\ U & -Q \end{pmatrix},
}
this transformation can be expressed as
\Eq{
(\bm{e}_x', \bm{e}_y') = (\bm{e}_x, \bm{e}_y) R
\then
P'= R^{-1} P R,
}
where $R$ is a two-dimensional rotation matrix, because the components of $P$ can be written
\Eq{
P_{ab}= 4\EXP{E_a E_b}- I \delta_{ab},\quad (a,b=x,y).
}

Let us introduce the complex null basis defined by
\Eq{
\bm{m}:= \frac{1}{\sqrt{2}}(\bm{e}_x + i \bm{e}_y),\quad
\b{\bm{m}}:= \frac{1}{\sqrt{2}}(\bm{e}_x - i \bm{e}_y).
}
Then, for the rotation of angle $\alpha$ represented by $R(\alpha)$, $\bm{m}$ transforms as
\Eq{
\bm{m}' = e^{i\alpha}\bm{m}.
}
Hence, the complex quantity $A$ by
\Eq{
A:= Q + iU = m^a m^b P_{ab},
}
transforms as
\Eq{
A' = e^{2i\alpha} A.
}
Thus, $A$ follows a simpler transformation law than $Q$ and $U$, but still depends on the choice of  the basis. This implies that we have to specify the basis in order to give a definite meaning to the values of $Q$ and $U$. We will give such a specification below.

\subsection{Polarisation Distribution Patterns}
\label{sec:pat}

Up to this point, we have considered a wave propagating only in one direction, whose polarisation can be described by the Stokes parameters $(I, Q, U)$. In real observations, this set of parameters is measured for photons of each direction, and the result is represented by three functions on the sky, $I(\bm{\theta}_\mathrm{obs})$, $Q(\bm{\theta}_\mathrm{obs})$ and $U(\bm{\theta}_\mathrm{obs})$, where $\bm{\theta}_\mathrm{obs}$ represents the position on the sky.  We define ${\bm{n}} := \bm{\theta}_{\mathrm{obs}} = (\theta_{\mathrm{obs}}, \varphi_{\mathrm{obs}})$. We often omit the subscript ``obs''  for brevity if no confusion occurs.

Let $\Theta$ denote the intensity (temperature) fluctuation around the sky average (`2.725K'):
$\Theta({\bm{n}}) := {\Delta I ({\bm{n}})}/{(4\bar{I})}$.
This distribution is expanded by spherical harmonic functions as
\Eq{
\Theta ({\bm{n}}) = \sum_{\ell m} \Theta_{\ell}^{m} Y_{\ell}^m ({\bm{n}}) \ . 
\label{equ:ThetaThetalm} 
}
Inversely, 
\Eq{
\Theta_\ell^m = \int d {\bm{n}} \Theta({\bm{n}}) Y_\ell^{m*}({\bm{n}}) \ . 
\label{equ:ThetalmTheta} 
}

Next, let us turn to the polarisation distribution. In order to define the Stokes parameters $Q(\bm{n})$ and $U(\bm{n})$ for each direction $\bm{n}$, we use the canonical orthonormal basis on the unit sphere with respect to the angular coordinates, $\{\bm{e}_\phi,\bm{e}_\theta\}$. Then, as in the previous subsection, we can define the polarisation tensor $P(\bm{n})$ on each direction $\bm{n}$  from $Q$ and $U$, which can be expressed in terms of the complex Stokes parameters on the sphere, ${}_{+2}A({\bm{n}})$ and ${}_{-2} A({\bm{n}})={}_{+2}\b A({\bm{n}})$, as
\Eq{ 
P_{ab} ({\bm{n}}) := {}_{+2}A({\bm{n}}) \bar{\bm{m}_a} \bar{\bm{m}_b} + {}_{-2} A({\bm{n}}) \bm{m}_a \bm{m}_b \ ,
\label{equ:Pm} 
}
where the subscripts $a$ and $b$ run over $\theta$ and $\varphi$ ($\theta \equiv 1$, $\varphi \equiv 2$). These complex Stokes parameters transform under the angular coordinate transformation induced by a rotation of the sphere represented by a 3-matrix $O$ as
\Eq{
{}_{\pm 2} A'({\bm{n}}) = e^{\pm 2i\alpha_O(\bm{n})} {}_{\pm 2}A(O^{-1}\bm{n}),
}
where $\alpha_O(\bm{n})$ is the rotation angle of the orthogonal frame induced by the transformation $O$. Functions on the unit sphere that transform in this way are called functions of spin-weight $\pm2$.  In general, these functions can be expanded in terms of spherical harmonics with spin-weight $\pm2$ as 
\Eq{ 
{}_{\pm 2} A({\bm{n}}) = \sum_{\ell m} {}_{\pm 2} A_\ell^m {}_{\pm 2} Y_\ell^m ({\bm{n}}) \ . 
\label{equ:AY} 
}
Here, the spin-weighted spherical harmonics are defined as\cite{1966JMP.....7..863N}\cite{1967JMP.....8.2155G}
\Eqr{
{}_s Y_\ell^m &=& \sqrt{\frac{(\ell-s)!}{(\ell+s)!}} \ \eth^s Y_\ell^m \ , \ \ \  0\leq s \leq \ell \ ; 
\label{equ:spinweightedpositive} \\
{}_s Y_\ell^m &=& \sqrt{\frac{(\ell+s)!}{(\ell-s)!}} \ (-1)^s\bar{\eth}^{-s} Y_\ell^m \ , \ \ \  -\ell \leq s \leq 0 \ ; 
\label{equ:spinweightednegative} \\
{}_s Y_\ell^m &=& 0 \ , \ \ \ \ell < |s| \ ; 
\label{equ:spinweightedzero} 
}
where
\Eqr{
\eth \eta &=& - (\sin \theta)^s \left( \frac{\partial}{\partial \theta} + \frac{i}{\sin \theta} \frac{\partial}{\partial \varphi} \right) \left\{ (\sin \theta )^{-s} \eta \right\} \ , 
\label{equ:eth} \\
\bar{\eth} \eta &=& - (\sin \theta)^{-s} \left( \frac{\partial}{\partial \theta} - \frac{i}{\sin \theta} \frac{\partial}{\partial \varphi} \right) \left\{ (\sin \theta )^{s} \eta \right\} \ , 
\label{equ:ethbar}
}
or to be explicitly by
\Eqr{
{}_s Y_{\ell}^m (\theta, \varphi) &=& 
\sqrt{ \frac{(\ell+m)!(\ell-m)!(2\ell+1)}{4\pi(\ell +s)!(\ell -s)!}  } \sin^{2\ell} \frac{\theta}{2} 
\nonumber \\
& & \cdot \sum_{r=0}^{\ell -s} \left( \begin{array}{c} \ell - s \\ r \end{array} \right) 
   \left(\begin{array}{c} \ell + s \\ r+s-m \end{array} \right) 
    (-1)^{\ell-r-s} e^{im\varphi} \cot^{2r+s-m} \frac{\theta}{2} \ . 
\label{equ:spinweightedcalc}
}
Note that ${}_0 Y_\ell^m = Y_\ell^m$. These functions transform under the rotation $O$ as
\Eq{
{}_s Y_{\ell}^m (O^{-1}\bm{n})=e^{-is \alpha_O(\bm{n})}
 {}_s Y_{\ell}^{m'} (\bm{n}) D^{m'm}_\ell(O),
}
where $D^{m m'}_\ell(O)$ is the $\ell$-th irreducible representation of the 3-dimensional rotation group. From this it follows that the expansion coefficients ${}_{\pm 2} A_\ell^m$ transform under the 3-rotation in the same way as the harmonic expansion coefficients for the temperature anisotropy $\Theta^m_\ell$:
\Eq{
{}_{\pm 2} A'{}_\ell^{m' }= D_\ell^{m'm}(O){}_{\pm 2} A_\ell^m.
}
Hence, we can define frame-independent harmonic expansion coefficients for the polarisation distribution by
\Eq{ 
E_\ell^m := \frac{1}{2}({}_{+2} A_\ell^m + {}_{-2} A_\ell^m ) \ , \quad
B_\ell^m := \frac{1}{2i}({}_{+2} A_\ell^m - {}_{-2} A_\ell^m ) \ . 
\label{equ:EBbyA} 
}
These represent the gradient (``E-mode'') and rotational (``B-mode'') components of the polarisation field, respectively.

\subsection{Power Spectra from Primordial Fluctuations}
\label{sec:primordial}

When the CMB temperature and polarisation anisotropies come only from primordial fluctuations, without any secondary effects such as gravitational lensing effects, the power spectra are defined in terms of the correlation functions as
\Eq{
\langle {X_1}_{\ell}^{m *} {X_2}_{\ell'}^{m'} \rangle
= C^{\mathrm{X_1} \mathrm{X_2}}_{\ell} \delta_{{\ell}{\ell}'} \delta_{mm'} \, ,
\label{equ:s2powerspectrum} 
}
or
\Eq{
C^{\mathrm{X}_1 \mathrm{X}_2}_{\ell} = \sum^{{\ell}}_{m 
   = -{\ell}} \frac{{X_1}_{\ell}^{m *} {X_2}_{\ell}^{m} }{2 {\ell} + 1} \ , 
 \label{equ:Cl} 
}
with ${X}_1,\ {X}_2 \in \{ {\Theta},\ {E},\ {B} \}$ and $\mathrm{X}_1,\ \mathrm{X}_2 \in \{ \mathrm{T},\ \mathrm{E},\ \mathrm{B} \}$. Here, the angle brackets represent an ensemble average over initial conditions; this average can be replaced in calculations with an average over space when the corresponding angular scale is much smaller than the observed region size.
If physics and the initial condition are invariant under a parity inversion, we have  $C^{\mathrm{TB}}_\ell =C^{\mathrm{EB}}_\ell = 0$.

\section{Gravitational Lensing Effects in the LTB model}
\label{sec:gl}

Inhomogeneous gravitational fields produce two effects on photon propagation. The first is a bending of its trajectory, and the second is the change of the photon energy in addition to the standard redshift by cosmic expansion. The latter is the so-called Sachs-Wolfe effect, which we do not consider in this article. Intuitively speaking, as far as CMB measurements by a fixed observer are concerned, the former---called `shear field effect'---can be further divided into two parts: ({\em{i}}) the change of the photon direction in the sky and ({\em{ii}}) the displacement of the intersection sphere of the past light cone and the last scattering surface in the direction perpendicular to this sphere. In order to give a definite meaning to this distinction, we need to introduce some reference FLRW model to define `unperturbed' photon trajectories and past light cones. However, this procedure introduces the gauge freedom corresponding to the mapping between the real universe and the reference model, and thus make that distinction obscure.  In fact, for the FLRW model with small perturbations, the displacement of the last scattering sphere can be set to be zero by an appropriate gauge choice, and in this gauge, the shear field effect can be represented only in terms of ({\em{i}}), namely, the `gravitational lensing effect'. In the local void model, it is not so certain whether the same argument holds when the non-linearity of inhomogeneities is large. In the present article, we simply assume that the shift of the last scattering point in the direction normal to the last scattering sphere can be set to zero by a gauge choice. 


Under this assumption, the gravitational lensing effect on the CMB anisotropy can be simply determined by the two-dimensional shift vector $\delta\bm{\theta}_\mathrm{obs}$ on the sky representing the difference between the observed direction of a photon and its initial direction on the last scattering sphere.

\begin{figure}[t]
\centering
\includegraphics[keepaspectratio=true,width=12cm]{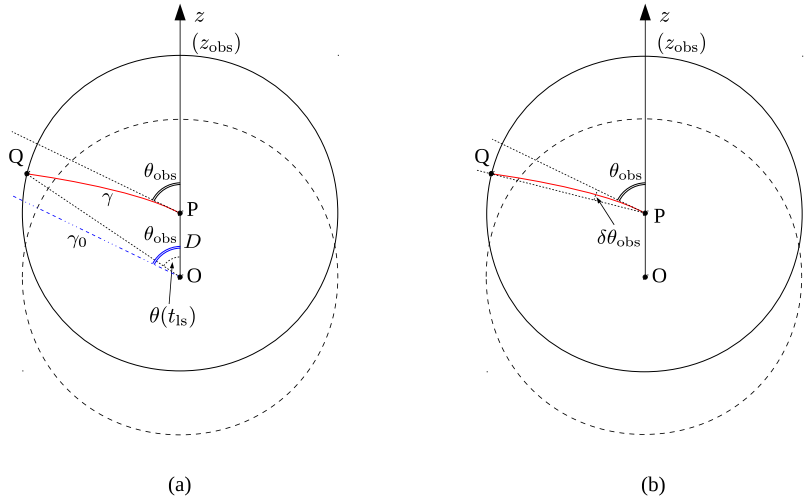}
\caption{
(a) 
Photon propagation in the LTB model. 
Each trajectory is contained in the unique two-plane passing through the center O, an off-center observer P, and the corresponding last scattering point Q. 
Without loss of generality, the two-plane with $\varphi = 0, \pi$ is selected as that unique two-plane.
The solid curve (red) represents the photon trajectory $\gamma$, while the dashed-dotted line (blue) represents the reference radial null geodesic $\gamma_0$. 
The solid circle represents the last scattering sphere for the off-center observer P, while the dashed circle represents that for the observer at the center O.
(b) 
The shift vector $\delta \bm{\theta}_\mathrm{obs}$.
Here, only the $\theta$ component $\delta {\theta}_\mathrm{obs}$ is nonzero because of the above selection of the two-plane.
}
\label{fig:fig1}
\end{figure}

\subsection{Null Geodesics in the LTB Model}
\label{sec:ge}

Thus, the investigation of the gravitational lensing effect of a local void on CMB is reduced to determine $\delta\bm{\theta}_\mathrm{obs}$ as a function of the photon direction. 
For that, we have to solve the null geodesic equation in the LTB model, whose metric can be written 
\Eq{
ds^2 = -dt^2 + S^2 d\chi^2 + r^2(d\theta^2 + \sin^2 \theta d\varphi^2).
\label{LTB:metric}
}
Here $r$ is a function of $t$ and $\chi$, and $S$ is written in terms of $r$ and the curvature function $k(\chi)$ as 
\Eq{
S=r'/(1-k(\chi)\chi^2)^{1/2}.
\label{curvaturefn:def}
}

In terms of the photon four-momentum $p^\mu=dx^\mu/d\lambda$ with affine parameter $\lambda$, the geodesic equation can be written as ${d p^\mu}/{d \lambda} = - \Gamma^\mu_{\nu\rho} p^\nu p^\rho$. 
Because of the spherical symmetry, this set of equations can be reduced to the coupled ODEs for $\omega$, $\mu$ and $p_\perp$ defined by 
\Eq{
p^t = \omega, \quad
p^\chi = \mu \omega /{S}, \quad
p_\perp^2 = \omega^2 (1-\mu^2),
}
where 
\Eq{
p_\perp := r \{( p^{\theta})^2 + (p^{\varphi})^2 \sin^2 \theta \}^{1/2}.
}
Note that $\mu$ represents the cosine of the angle between the propagation direction of the photon and the radial direction of the photon position from the center. 

Without loss of generality, we can assume that the photon propagates on the 2-plane with $\varphi=0,\pi$, and therefore $p^\varphi=0$. 
Then, the geodesic equations are reduced to the set of four ODEs for $\omega(t)$, $\chi(t)$, $\theta(t)$, and $\mu(t)$:
\Eqrsub{
\frac{d}{dt}\ln\omega &=& - \frac{\dot{S}}{S} \mu^2 - (1-\mu^2)\frac{\dot{r}}{r} \ , 
\label{equ:dlnomegadt} \\
\frac{d\chi}{dt} &=& \frac{\mu}{S} \ , 
\label{equ:dchidt} \\
\frac{d\theta}{dt} &=& \pm\frac{\sqrt{1-\mu^2}}{r} \ , 
\label{equ:dthetadt} \\
\frac{1}{1-\mu^2} \frac{d\mu}{dt} &=& \frac{\xi}{r} 
+ \mu \left( \frac{\dot{r}}{r} - \frac{\dot{S}}{S} \right) 
\label{equ:dmudt} \ , 
}
where $\xi=(1-k(\chi)\chi^2)^{1/2}$.
Note that \eqref{equ:dchidt} and \eqref{equ:dmudt} form a closed set of ODEs for $\chi(t)$ and $\mu(t)$, and  $\omega$ and $\theta$ can be determined from each solution for $\chi(t)$ and $\mu(t)$ by simply integrating knowing functions, using (\ref{equ:dlnomegadt}) and (\ref{equ:dthetadt}).

We illustrate our setup in Figure \ref{fig:fig1}a.
Let O denote the center of the void and P the position of the observer in the comoving coordinate chart. We will take the $z$-axis to run through O and P in order. Let us consider a radial null ray $\gamma_0$ with the angle $\theta_\mathrm{obs}$ relative to OP and a null ray $\gamma$ that passes through P with $\mu= - \cos \theta_\mathrm{obs}$. Then, if $D=\overline{\mathrm{OP}}$ is small, $\gamma$ stays close to $\gamma_0$ until the last scattering surface at $t=t_\mathrm{ls}$. Hence, its behaviour can be determined by solving the linear perturbation equation obtained from the above ODEs for $\chi$ (\ref{equ:dchidt}) and $\mu$ (\ref{equ:dmudt}).

One subtle point of this approach is that $\mu$ cannot be treated perturbatively, because $\mu= -1$ for $\gamma_0$, but $\mu$ changes largely for $\gamma$ around P. This difficulty can be avoided by using the variables $b$ and $c$ defined by
\Eq{
b=\chi\sqrt{1-\mu^2} \ , \ \ \ c=\chi \mu,
\label{equ:bc} 
}
in stead of $\chi$ and $\mu$. In fact, the geodesic equations can be written in terms of $b$ and $c$ as
\Eqrsub{ 
\dot{b} &=& \xi \alpha b c - \beta b c^2 \ , \label{equ:dotb} \\
\dot{c} &=& \frac{1}{S} - \xi \alpha b^2 + \beta b^2 c \ , \label{equ:dotc}
}
where $\alpha(t,\chi)$ and $\beta(t, \chi)$ are
\Eq{
\alpha = \frac{1}{\chi^2} \left( \frac{1}{r'} - \frac{\chi}{r} \right) \ ,\quad
\beta = \frac{1}{\chi^2} \left( \frac{\dot{r}}{r} - \frac{\dot{S}}{S} \right) \ . 
\label{equ:alphabeta} 
}
Note that these are regular at $\chi=0$ in general and vanish for a spatially homogeneous background. From these, it follows
\Eq{
b\frac{d}{dt}\left(\frac{c}{b}\right) = \frac{1}{S} + \chi^2(-\alpha \xi + \beta c) \ .
\label{equ:bddtcb} 
}

The perturbation equation of the geodesic equation up to the first order in $b$ now reads
\Eqrsub{
\dot{b} &=& -X b \ , 
\label{equ:dotdeltab} \\
\dot{\delta c} &=& \frac{S'}{S^2} \delta c \ , 
}
where $X(t)$ is the function on the central past light cone $\chi=\chi_0(t)$, 
\Eq{
X(t) = ( \chi \xi \alpha + \chi^2 \beta )_{\chi=\chi_0(t)} \ . 
\label{equ:X}
}
These can be easily solved to yield
\Eq{
b(t) = b(t_0) e^{Y(t,t_0)} \ , 
\label{equ:bt} 
}
where
\Eq{
Y(a,b)= \int_a^b dt X(t) \ . 
\label{equ:Y} 
}
%

\subsection{The Shift Vector $\delta\bm{\theta}_\mathrm{obs}$}
\label{sec:rel}

In order to estimate the shift vector $\delta\bm{\theta}_\mathrm{obs}$ (Fig.~\ref{fig:fig1}b), we need to calculate $\theta(t)$.
We can assume that $\theta(t_0)=0$ without loss of generality.
In addition, assuming that $\theta(t)$ monotonically decreases from $t_1$ to $t_0$, we choose the minus sign in (\ref{equ:dthetadt}).
Then, $\theta(t)$ can be written
\Eqr{
\theta(t_1) 
&=& - \int_{t_0}^{t_1} dt \frac{b}{\chi r} 
= - \int_{t_0}^{t_1} \frac{dt}{b} \frac{\chi/r}{1+(c/b)^2} \nonumber \\
&=& - \int_{t_0}^{t_1} \frac{(c/b) \dot{\ } dt}{1 + (c/b)^2} \frac{\chi/r}{1/S - \chi^2 \xi \alpha + \chi^2 c \beta} \nonumber \\
&=& - \int_{(c/b)(t_0)}^{(c/b)(t_1)} \frac{d(c/b)}{1 + (c/b)^2} - \int_{t_0}^{t_1} \frac{(c/b)\dot{\ } dt}{1 + (c/b)^2} \left\{ \frac{\chi/r}{1/S - \chi^2 \xi \alpha + \chi^2 c \beta} -1 \right\} \nonumber \\
&=& \theta_\mathrm{obs} - \tan^{-1} \left. \frac{b}{\chi} \right|_{t_1} - \int_{t_0}^{t_1} dt b \left\{ \frac{1 - \xi}{\chi^2}\frac{\chi}{r} + \chi \beta \right\} \ , 
\label{equ:thetat1} 
}
where we have set $c = - \chi$ that holds on the radial null ray.
Further, the integral $I$ in the last expression can be deformed with the help of a partial integration as
\Eqr{
I &=& \int_{t_0}^{t_1} dt b 
  \left\{ \frac{1 - \xi}{\chi^2}\frac{\chi}{r} + \chi \beta \right\} 
  \nonumber \\
&=& b(t_0) \int_{t_0}^{t_1} dt \left\{ e^{Y(t,t_0)} 
  \left( \frac{1-\xi}{\chi^2}\frac{\chi}{r} - \xi \alpha \right) 
   - \frac{1}{\chi} \frac{d}{dt} (e^{Y(t,t_0)} - 1) \right\} 
   \nonumber \\
&=& \left[ - \frac{b(t_0)}{\chi} (e^{Y(t,t_0)} - 1) \right]_{t_0}^{t_1} 
  + b(t_0) \int_{t_0}^{t_1} \frac{dt}{\chi^2} 
   \left\{ e^{Y(t,t_0)}\left(\frac{\chi}{r} - \frac{1}{S} \right) 
   - \dot{\chi} (e^{Y(t,t_0)} - 1) \right\} 
   \nonumber \\
&=& - \frac{b(t_1) - b(t_0)}{\chi(t_1)} 
 + b(t_0) \int_{t_0}^{t_1} \frac{dt}{\chi^2} 
  \left( - \frac{1}{S} + \frac{\chi}{r} e^{Y(t,t_0)} \right) \ . 
\label{equ:Integral}
}
Hence, we obtain
\Eq{
\theta(t_\mathrm{ls}) = \theta_\mathrm{obs} - \frac{b(t_0)}{\chi_\mathrm{ls}}
   - b(t_0) \int_{t_\mathrm{ls}}^{t_0} \frac{dt}{\chi^2} 
    \left( \frac{1}{S} - \frac{\chi}{r} e^{Y(t,t_0)} \right) \ . 
 \label{equ:thetatls} 
}
Thus, by eliminating the part that survives in the spatially homogeneous limit, we find that the shift in the angular direction of the null geodesic due to the void shear with respect to an observer P at distance $D$ from the symmetry center O is given by
\Eq{
\delta\theta_\mathrm{obs} = - D \sin \theta_\mathrm{obs}  
 \int_{t_\mathrm{ls}}^{t_0} \frac{dt}{\chi^2} 
 \left( \frac{\chi}{r}(1- \xi) + \frac{1}{S} - \frac{\chi}{r} e^{Y(t,t_0)} \right) \ ,
\label{equ:deltathetaobs} 
}
where $\theta_\mathrm{obs}$ is the angle of the null geodesic direction with respect to the observer direction OP.
This angular shift vector is used in Section \ref{sec:result}.

\section{CMB Polarisation in the LTB model}
\label{sec:polLTB}

\subsection{Flux Intensity Tensor}
\label{sec:fluxtensor}

In the Lorentz gauge, the free electromagnetic potential $A_\mu$ can be written in terms of the creation and annihilation operators as
\Eq{
A_\mu(x) = \int \frac{d^3 \bm{k}}{(2 \pi)^3} \frac{1}{2 \omega} 
\sum_p ( e_{p \mu} (\bm{k}) a_p(\bm{k}) e^{ik \cdot x} + e_{p \mu}^* (\bm{k}) a_p (\bm{k})^\dag e^{-ik \cdot x} ) \ ,
\label{equ:Amu} 
}
where $e_{p \mu}(\bm{k})$ is the polarisation basis satisfying
\Eq{
e_{p}^\mu e_{q \mu}^* = \delta_{p q} \ , \ \ \ k^\mu e_{p \mu} = 0 \ .
 \label{equ:polbasiscond1} 
}
Note that the addition of vectors proportional to $k^\mu$ to $e_p^\mu$ has no physical significance, because it corresponds to a gauge transformation and produces no physical effect in the exact quantum formulation.
$a_p$ and $a_p^\dag$ satisfy the standard relativistic commutation relations
\Eq{
[a_p(\bm{k}) , a_q (\bm{k}') ] = 0 \ , \ \ \ [a_p(\bm{k}), a_q(\bm{k}')^\dag ] = (2 \pi)^3 2 \omega \delta_{p q} \delta^3(\bm{k} - \bm{k}') \ . 
\label{equ:commutation} 
}
The electric field $\bm{E}$ and the magnetic field $\bm{B}$ are 
\Eqrsub{
\bm{E} &=& \int \frac{d^3 \bm{k}}{(2 \pi)^3} \frac{1}{2} 
\sum_{p} ( \bm{\epsilon}_p(\bm{k}) a_p (\bm{k}) e^{ik \cdot x} + \bm{\epsilon}_p^* (\bm{k}) a_p(\bm{k})^\dag e^{-ik \cdot x} ) \ , \label{equ:ele} \\
\bm{B} &=& \int \frac{d^3 \bm{k}}{(2 \pi)^3} \frac{1}{2 \omega} \bm{k} \times
\sum_{p} ( \bm{\epsilon}_p(\bm{k}) a_p (\bm{k}) e^{ik \cdot x} + \bm{\epsilon}_p^* (\bm{k}) a_p(\bm{k})^\dag e^{-ik \cdot x} ) \ , \label{equ:mag} 
}
where
\Eq{
\epsilon_{p j} = e_{p j} - \frac{k_j}{\omega} e_{p 0} \ ,
 \label{equ:polbasis} 
}
which satisfies
\Eq{
\bm{k} \cdot \bm{\epsilon}_p = 0 \ , \ \ \ \bm{\epsilon}_p^* \cdot \bm{\epsilon}_q = \delta_{pq} \ . 
\label{equ:polbasiscond2} 
}

Now, let us define the measured components of the electric field, $\mathcal{E}_p$, in terms of the sensitivity function $W(\bm{x})$ and the detector polarisation basis $\bm{\epsilon}^\mathrm{o}_p$ as 
\Eq{
\mathcal{E}_p = \int d^3 \bm{x} W(\bm{x}) \bm{\epsilon}_p^\mathrm{o} \cdot \bm{E} (t_0 , \bm{x}) \ . 
\label{equ:elemeasured} 
}
Then, for the free field, $\mathcal{E}_p$ can be expressed as
\Eq{
\mathcal{E}_p = \int \frac{d^3 \bm{k}}{(2 \pi)^3} \frac{1}{2} 
\left[ \sum_q a_q (\bm{k}) (\bm{\epsilon}^\mathrm{o}_p \cdot \bm{\epsilon}_q (\bm{k}) ) \hat{W}(\bm{k}) e^{-i \omega t} + \mathrm{c.c.} \right] \ , 
\label{equ:elemeasured2} 
}
where
\Eq{
\hat{W}(\bm{k}) = \int d^3 \bm{x} e^{i \bm{k} \cdot \bm{x} } W(\bm{x}) \ . 
\label{equ:hatW} 
}

Now, assume that
\Eq{
\langle a_p(\bm{k}) a_q (\bm{k}') \rangle = 0 \ , \ \ \  
\langle a_p (\bm{k})^\dag a_q (\bm{k}') \rangle = 2 (2 \pi)^3 \rho_{p q} (\bm{k}) \delta^3 (\bm{k} - \bm{k}') \ .
\label{equ:correlation}
}
Then, the observed correlation of the electric fields can be written
\Eqrsub{
\langle : \mathcal{E}_p \mathcal{E}_q : \rangle &=& {\epsilon^\mathrm{o}_p}^i {\epsilon^\mathrm{o}_q}^j \int \frac{d^3 \bm{k}}{(2 \pi)^3} |\hat{W} (\bm{k})|^2 \rho_{(i j)} (\bm{k}) \ , \label{equ:elecorr1} \\
\langle : \mathcal{E}_p \tilde{\mathcal{E}}_q : \rangle &=& {\epsilon^\mathrm{o}_p}^i {\epsilon^\mathrm{o}_q}^j \int \frac{d^3 \bm{k}}{(2 \pi)^3} |\hat{W} (\bm{k})|^2 (-i) \rho_{[i j]} (\bm{k}) \ , \label{equ:elecorr2}
}
where
\Eq{ 
\rho_{i j} (\bm{k}) = \sum_{p,q} \epsilon_{p i}^* (\bm{k}) \epsilon_{q j} (\bm{k}) \rho_{p q}(\bm{k}) \ ,
\label{equ:rhoij} 
}
and $\tilde{\mathcal{E}}_p$ is obtained from $\mathcal{E}_p$ by advancing the phase $\omega t$ by $\pi/2$ for each mode.
Note that the Stokes parameters are
\Eqr{ 
&& I = \sum_p {\epsilon^\mathrm{o}_p}^i {\epsilon^\mathrm{o}_p}^j \rho_{(ij)} \ ,\quad
   Q = ({\epsilon^\mathrm{o}_1}^i {\epsilon^\mathrm{o}_1}^j - {\epsilon^\mathrm{o}_2}^i {\epsilon^\mathrm{o}_2}^j ) \rho_{(i j)} \ , 
   \notag\\
&& U = 2 {\epsilon^\mathrm{o}_1}^i {\epsilon^\mathrm{o}_2}^j \rho_{(i j)} \ , \quad
   V = (- 2 i ) {\epsilon^\mathrm{o}_1}^i {\epsilon^\mathrm{o}_2}^j \rho_{[i j]} \ .
\label{equ:stokes} 
}
Thus, $\rho_{i j} (\bm{k})$ provides a polarisation-basis-independent description of the radiation field polarisation and intensity.
We call $\rho_{i j}$ and $\rho_{p q}$ the flux density tensor and the flux polarisation matrix, respectively.

\subsection{Polarised Boltzmann Equation}
\label{sec:polboltzmann}

In a curved spacetime, the above mode functions for the expansion of the electromagnetic fields should be replaced by corresponding vector fields satisfying
\Eq{
\nabla^\mu F_{\mu \nu} = 0 \ . 
\label{equ:Fcond} 
}
Accordingly, it is rather difficult to treat wavefunctions and polarisation vectors independently.
However, such a treatment is allowed for modes for which the WKB approximation is good.
For such modes, the mode function can be written
\Eq{ 
A_\mu(x) = a_\mu (x) e^{i S(x)} \ , 
\label{equ:Amu2} 
}
where for $k := \nabla S$, $a_\mu(x)$ and $S(x)$ satisfy
\Eqrsub{ 
&& k := \nabla S \ \Longrightarrow \ k \cdot k \approx 0 \ \Longrightarrow \ \nabla_k k \approx 0 \ ,\\
\label{equ:aScond1} 
&& \nabla_k a_\mu = - \frac{1}{2} \Box S a_\mu \approx 0 \ . 
\label{equ:aScond2} 
}
Hence, by generalising the polarisation basis to spacetime-dependent vectors $\epsilon^\mu(k , x)$ such that
\Eq{ 
\nabla_k \epsilon^\mu (k, x) = 0 \ , \ \ \ k_\mu \epsilon^\mu (k, x) = 0 \ , 
\label{equ:polbasiscond3} 
}
the flux density tensor $\rho_{i j}$ can be generalised to 
\Eq{ 
\rho^{\mu \nu} = \sum_{p , q} \epsilon_p^{\mu *} \epsilon_q^\nu \rho_{pq} \ . 
\label{equ:rhomunu} 
}
This tensor is independent of the polarisation basis and satisfies the generalised Boltzmann equation
\Eq{ 
\left(\frac{k^\sigma}{k^0} \nabla_\sigma + \frac{\dot{k^i}}{k^0} \partial_{k^i} \right) \rho^{\mu \nu} (x, \bm{k}) = C^{\mu \nu} (\rho) \ , 
\label{equ:polboltzmann} 
}
where $C^{\mu \nu}$ is the collision term.

\subsection{The Initial Condition at the Last Scattering Surface}
\label{sec:initcond}

On the last scattering surface, up to the linear order in perturbations, $\delta \rho_{\mu \nu}$  can be Fourier decomposed into the contribution of each perturbation mode with the wave vector $\bm{K}$ as
\Eq{ 
\label{equ:deltarhomunu} 
\delta \rho_{\mu \nu} (t_\mathrm{ls}, \bm{x}, \bm{k}) = \int d^3 \bm{K} e^{i \bm{K} \cdot \bm{x}} \rho^{(1)}_{\mu \nu} (\bm{K}; \bm{k}) \ . 
}
In this situation, it is customary to adopt the following polarisation basis: 
\Eq{ 
\label{equ:polbasisinit} 
\bm{\epsilon}_p(k) = \frac{1}{a} \hat{\bm{\epsilon}}_p(k) \ ; 
}
\Eqrsub{
\hat{\bm{\epsilon}}_1 &=& \frac{1}{\sqrt{1 - (\hat{\bm{k}} \cdot \hat{\bm{K}})^2}} \left( \hat{\bm{K}} - (\hat{\bm{k}} \cdot \hat{\bm{K}}) \hat{\bm{k}} \right) \ , \label{equ:polbasisinit1} \\
\hat{\bm{\epsilon}}_2 &=& \frac{1}{\sqrt{1 - (\hat{\bm{k}} \cdot \hat{\bm{K}})^2}} \hat{\bm{K}} \times \hat{\bm{k}} \ , \label{equ:polbasisinit2} 
}
where $a$ is the scale factor\footnote{The early universe can be treated as the FLRW universe plus a perturbation as described below.} and
\Eq{ 
\label{equ:kKnormalisation} 
\hat{\bm{k}} = \frac{\bm{k}}{|\bm{k}|} \ , \ \ \ \hat{\bm{K}} = \frac{\bm{K}}{|\bm{K}|} \ . 
}
In the spherical coordinates corresponding to the Cartesian coordinates in which 
\Eq{ 
\label{equ:Kk} 
\hat{\bm{K}} = (0,0,1) \ , \ \ \ \hat{\bm{k}} = ( \sin \theta \cos \varphi, \sin \theta \sin \varphi, \cos \theta) \ , 
}
the polarisation basis has the following components:
\Eqrsub{ 
\hat{\bm{\epsilon}}_1 &=& (\cos \theta \cos \varphi, \cos \theta \sin \varphi, - \sin \theta) \ , \label{equ:polbasisinit10} \\ 
\hat{\bm{\epsilon}}_2 &=& (- \sin \varphi, \cos \varphi, 0 ) \ . \label{equ:polbasisinit20}
}
For this choice of the polarisation basis, because all vector-like quantities for a perturbation is parallel to $\hat{\bm{K}}$, $\rho^{(1)}_{p q}$ defined by
\Eq{ 
\label{equ:rho1pq} 
\rho^{(1)}_{p q} = \epsilon_p^{\mu *} \epsilon_q^\nu \rho^{(1)}_{\mu \nu} (\bm{K}; \bm{k}) 
}
depends only on $\omega$ and $\mu = \cos \theta = \hat{\bm{K}} \cdot \hat{\bm{k}}$:
\Eq{ 
\label{equ:rho1pq0} 
\rho^{(1)}_{p q} =  \rho^{(1)}_{p q} ( \bm{K}; \omega, \mu) \ . 
}

Based on this setup, we can solve the Boltzmann equation (\ref{equ:polboltzmann}). This can be done following the standard procedure used in the case of the FLRW universe\cite{1984ApJ...285L..45B}\cite{1996AnPhy.246...49K}, and finally we obtain the initial condition at the last scattering surface.

\begin{figure}[t]
\centering
\includegraphics[keepaspectratio=true,width=12cm]{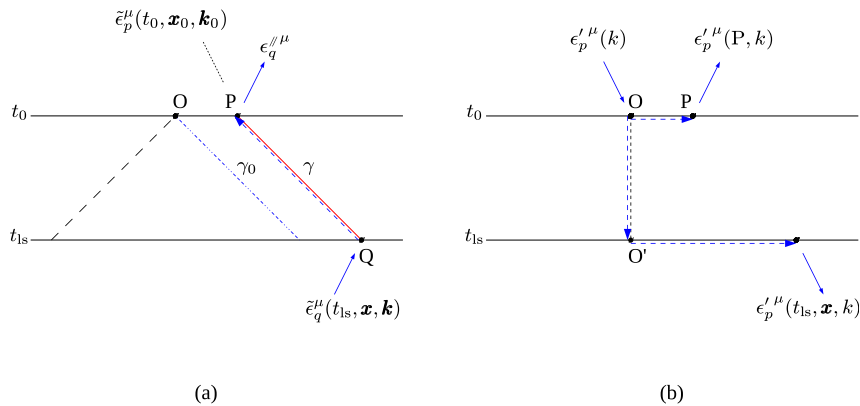}
\caption{
(a) Propagation of the polarisation basis in a general perturbed LTB spacetime.
The polarisation basis $\tilde{\epsilon}_q (t_\mathrm{ls}, \boldsymbol{x}(x_0, k_0), \boldsymbol{k}(x_0, k_0))$ at Q$(k_0)$ on the last scattering surface $t=t_\mathrm{ls}$ is parallelly propagated along the null geodesic $\gamma(\mathrm{P}, k_0)$ to P on the present-time hypersurface $t=t_0$ and named ${\epsilon^{\scriptscriptstyle /\!/}_q}$, which should be compared to the polarisation basis $\tilde{\epsilon}_p (t_0, \bm{x}_0, \bm{k}_0)$ at P. 
(b) The global polarisation basis in the unperturbed LTB background.
The polarisation basis ${{\epsilon}'_p} (\hat{\bm{z}}, k)$ at the present void center O, together with $k$, is transported to an arbitrary point P on the hypersurface $t=t_0$ along the radial line $\Omega = \text{const.}$ parallelly with respect to $g_{ij}(t_0, x)$ and becomes ${{\epsilon}'_p} (\mathrm{P},k)$.
Meanwhile, ${{\epsilon}'_p} (\hat{\bm{z}}, k)$ at O is transported back to the symmetry center O' at $t = t_\mathrm{ls}$ along the timelike path corresponding to the symmetry center to define the basis at O', which is consecutively extended to an arbitrary point on the last scattering surface $t=t_\mathrm{ls}$ by the parallel transport with respect to $g_{ij}(t_\mathrm{ls}, x)$ and becomes ${{\epsilon}'_p} (t_\mathrm{ls}, \bm{x}, k)$.
}
\label{fig:fig2}
\end{figure}

\subsection{Propagation After Last Scattering}
\label{sec:propagation}

Let us work in the synchronous gauge in which
\Eq{ 
\label{equ:synchro} 
ds^2 = -dt^2+ \tilde{g}_{ij} (t,x) dx^i dx^j \ , 
}
where $\tilde{g}_{ij}$ approaches a spatially homogeneous and isotropic metric $g_{ij}$ in the early universe:
\Eq{ 
\label{equ:tildegij} 
\tilde{g}_{ij} = g_{ij} + \delta g_{ij} \ . 
}
We take the constant-time surfaces so that the last scattering surface is represented by $t=t_\mathrm{ls}$.
Note that the LTB spacetime $g'_{\mu \nu}$ also belongs to this class:
\Eq{ 
\label{equ:gprimeij} 
g'_{ij} = g_{ij} + \delta_\mathrm{LTB} g_{ij} \ .
}
In most part, we take the spatially flat FLRW solution as the background $g_{\mu \nu}$.

Now, we consider the measurement of the CMB polarisation at the spacetime point $\mathrm{P}=(t_0, \bm{x}_0)$. The null geodesic $\gamma(\mathrm{P}, k_0)$ passing through P with the four-momentum $k_0 = (\omega(\bm{k}_0), \bm{k}_0)$ intersects with the last scattering surface at the point $\mathrm{Q}(k_0)$ with the space coordinates $\bm{x} = \bm{x}(x_0,k_0)$ and the four-momentum $k = k(x_0, k_0)$. Let us represent polarisation bases at each point on the hypersurfaces $t=t_0$ and $t=t_\mathrm{ls}$ as $\tilde{\epsilon}_p^\mu(t_0, \bm{x}, \bm{k})$ and $\tilde{\epsilon}_p^\mu(t_\mathrm{ls}, \bm{x}, \bm{k})$, respectively. Then, the flux polarisation matrix at P is expressed in terms of the corresponding quantities on the last scattering surface as
\Eq{ 
\label{equ:tilderho} 
\tilde{\rho} (\mathrm{P}, \bm{k}_0) = \tilde{C} \tilde{\rho} (t_\mathrm{ls}, \bm{x}, \bm{k}) \tilde{C}^\dag \ , 
}
where $\bm{x} = \bm{x} (x_0, k_0)$ and $\bm{k} = \bm{k} (x_0, k_0)$ are understood. The matrix $\tilde{C}$ is defined as follows. First, we parallelly propagate the polarisation basis $\epsilon_q (t_\mathrm{ls}, \bm{x}(x_0, k_0), \bm{k}(x_0, k_0))$ at $\mathrm{Q}(k_0)$ on the last scattering surface along the null geodesic $\gamma(\mathrm{P}, k_0)$ to P (Fig.~\ref{fig:fig2}a). Let $\epsilon^\para_q$ denote this basis at P. Then,
\Eq{ 
\label{equ:tildeC} 
\tilde{C}_{pq} = \tilde{\epsilon}_p (t_0, \bm{x}_0, \bm{k}_0) \cdot {\epsilon^\para_q}^* \ . 
}

When the universe is well-described by the FLRW model at and before the last scattering, $\tilde{\rho} (t_\mathrm{ls}, \bm{x}, \bm{k})$ is approximately isotropic:
\Eq{ 
\label{equ:tilderhopq} 
\tilde{\rho}_{pq}(t_\mathrm{ls}, \bm{x}, \bm{k}) = \frac{1}{2} \delta_{pq} I \left( \frac{\omega}{T_\mathrm{ls}} \right) + \delta \rho_{pq} (t_\mathrm{ls}, \bm{x}, \bm{k}) \ . 
}
Then, from the identity
\Eq{ 
\label{equ:tildeCid} 
\sum_p \tilde{C}_{pq} \tilde{C}_{pr}^* = \delta_{qr} \ , 
}
we obtain
\Eq{ 
\label{equ:tilderhopqP} 
\tilde{\rho}_{pq} (\mathrm{P}, \bm{k}_0) = \frac{1}{2} \delta_{p q} I \left( \frac{\omega}{T_\mathrm{ls}}\right) + \tilde{C}_{pr} \tilde{C}_{qs}^* \delta \rho_{rs} (t_\mathrm{ls}, \bm{x}, \bm{k}) \ . 
}
The first term on the right-hand side of this equation is only relevant to the temperature and the polarisation comes only from the second term. In this formulation, it is understood that $\delta \rho_{rs} (t_\mathrm{ls}, \bm{x}, \bm{k})$ is for a polarisation basis that does not depend on the mode wave number.  

Now, we show that $\tilde{C}_{pq}$ is close to the unit matrix in an appropriate global polarisation basis(Fig.~\ref{fig:fig2}b). First, we define the polarisation basis at the center O of the void at present as
\Eqrsub{ 
\bm{\epsilon}'_1(\hat{\bm{z}}, k) &=& \frac{1}{\sqrt{1- (\hat{\bm{z}} \cdot \hat{\bm{k}})^2}} \left(\hat{\bm{z}} - (\hat{\bm{k}} \cdot \hat{\bm{z}} ) \hat{\bm{k}} \right) \ , \label{equ:epsilonprime1} \\
\bm{\epsilon}'_2(\hat{\bm{z}}, k) &=& \frac{1}{\sqrt{1- (\hat{\bm{z}} \cdot \hat{\bm{k}})^2}} \hat{\bm{z}} \times \hat{\bm{k}} \ , \label{equ:epsilonprime2}
}
where $\hat{\bm{z}}$ is a unit spacelike vector and $\hat{\bm{k}} = \bm{k}/|\bm{k}|$ as before. Later, $\hat{\bm{z}}$ is taken to be the direction of the observer from O. Next, we transport this pair and $k^i$ to an arbitrary point P on the hypersurface $t=t_0$ along each radial line  parallelly with respect to the space metric $g_{ij} (t_0, x)$ to define the polarisation basis at P with respect to $\bm{k}$, ${\epsilon'_p}^\mu (\mathrm{P}, k)$. In this way, we can define a polarisation basis everywhere on the hypersurface $t=t_0$.

Next, we parallelly transport the basis at O back to the symmetry center O' at $t=t_\mathrm{ls}$ along the timelike path corresponding to the symmetry center to define the basis at O'.
Then, extend it to an arbitrary point on the last scattering surface $t=t_\mathrm{ls}$ as on the hypersurface $t=t_0$.
When the spacetime is exactly spherically symmetric, we can easily relate this basis to the mode-dependent basis introduced in the previous section, if we neglect the small spatial inhomogeneity of the LTB universe at $t=t_\mathrm{ls}$.
Because the matrix $\tilde{C}_{pq}$ that we are calculating is multiplied by a perturbation $\delta \rho_{pq}$ in (\ref{equ:tilderhopqP}), and we know $\delta \rho_{pq}$ only in the first order w.r.t.\ perturbations at last scattering, we can neglect the small inhomogeneity of the LTB universe at the last scattering in calculating $\tilde{C}_{pq}$.

Note that when the polarisation basis $\bm{\epsilon}_p(\bm{k})$ is given for a background space metric $g_{ij}$, we can uniquely determine the corresponding basis $\tilde{\bm{\epsilon}}_p(\bm{k})$ for $\tilde{g}_{ij} = g_{ij} + \delta g_{ij}$ by the requirement
\Eq{ 
\label{equ:polbasiscond4} 
\tilde{\bm{\epsilon}}_p \cdot \tilde{\bm{\epsilon}}_q = \delta_{pq} \ , \ \ \ \tilde{\bm{\epsilon}}_p \cdot \bm{k} = 0 
}
as
\Eq{ 
\label{equ:deltaepsilon} 
\delta \epsilon_{pj} = - \frac{1}{2} h_{j \ell} \epsilon_{pm} \delta g^{\ell m} \ , 
}
where
\Eq{ 
\label{equ:hij} 
h_{ij} = g_{ij} - \hat{{k}}_i \hat{{k}}_j \ . 
}
Hence, in order to determine the matrix $\tilde{C}_{pq}$, we only have to calculate the propagation of the polarisation basis for the background LTB universe neglecting small inhomogeneities.

Now, we show that the transfer matrix $\tilde{C}_{pq}$ for the polarisation basis can be well approximated by the unit matrix for the global polarisation basis introduced above. Because we are only interested in the shear effect of the LTB geometry, we can calculate this matrix by parallelly propagating the unperturbed basis ${\epsilon'_p}^\mu$ in the background LTB universe neglecting additional perturbations. Now, let us consider a null geodesic $\gamma_0$ passing through O at $t = t_0$ and a null geodesic $\gamma$ passing through an off-center observer at P close to O.
Then, it is easy to confirm that for the null vector $k$ parallel to $\gamma_0$, the corresponding polarisation basis ${\epsilon'_p}^\mu(t, k) $ at $t=t_0$ and $t=t_\mathrm{ls}$ defined above are parallelly related.

Let ${\epsilon''_p}^\mu(t, z) $ be a polarisation basis parallelly propagated along a family of null geodesics with the parameter $z$ close to $\gamma_0$ among which $\gamma$ is contained.
Then, along $\gamma$, we obtain
\Eq{ 
\label{equ:nablaepsilon2} 
\nabla_k [ \epsilon'_q \cdot ( \nabla_Z \epsilon''_p)] = \epsilon'_q \cdot R(k, Z) \epsilon'_p \ , 
}
where $\nabla_Z$ refers to the covariant derivative along the deviation vector $Z$.
This can be written
\Eq{ 
\label{equ:ddtepsilonnablaepsilon} 
\frac{d}{dt} ( \epsilon'_q \cdot \nabla_{\delta x} \epsilon''_p ) = {\epsilon'_q}^j {\epsilon'_p}^\ell R_{j \ell \mu m} \left(\frac{k^\mu}{\omega} \right) \delta x^m \ ,
}
where $\delta x$ is the short distance along $Z$.
Here, the polarisation basis $\bm{\epsilon}'_p$ on $\gamma$ can be written explicitly in the spherical coordinates for the LTB metric as
\Eq{ 
\label{equ:epsilonprime10} 
\bm{\epsilon}'_1 = - \frac{1}{r} \partial_\theta \ , \ \ \
\bm{\epsilon}'_2 = \frac{1}{r \sin \theta} \partial_\varphi \ . 
}
Further, the non-vanishing components of the Riemann tensor for the LTB metric are
\Eqrsub{
& & R_{t\chi t \chi} = -S \ddot{S} \ , \ \ \ {R_{t A t}}^B = -\frac{\ddot{r}}{r} \delta_A^B \ , \ \ \ {R_{t A \chi}}^B = \frac{S}{r} \left(\frac{r'}{S} \right)^{\bm{\cdot}} \delta_A^B \ , \label{equ:riemannt} \\
& & {R_{\chi A \chi}}^B = \frac{S \dot{S} \dot{r}}{r} - \frac{S}{r} \left(\frac{r'}{S} \right)' \ , \label{equ:riemannchi} \\
& & {R_{AB}}^{CD} = \frac{1}{r^2} \left\{ \dot{r}^2 + 1 - \left(\frac{r'}{S} \right)^2 \right\} ( \delta_A^C \delta_B^D - \delta_A^D \delta_B^C) \ . \label{equ:riemannA}
}
From these, it follows that the right-hand side of (\ref{equ:ddtepsilonnablaepsilon}) vanishes. Further, from the structure of the non-vanishing components of the Christoffel symbol for the LTB metric,
\Eqrsub{
& & \Gamma^t_{\chi \chi} = S \dot{S} \ , \ \ \ \Gamma^t_{AB} = r \dot{r} \gamma_{AB} \ , \label{equ:christt} \\
& & \Gamma^\chi_{t \chi} = \frac{\dot{S}}{S} \ , \ \ \ \Gamma^A_{tB} = \frac{\dot{r}}{r} \delta^A_B \ , \label{equ:christchi} \\
& & \Gamma^\chi_{\chi \chi} = \frac{S'}{S} \ , \ \ \ \Gamma^\chi_{AB} = - \frac{r r'}{S^2} \gamma_{AB} \ , \ \ \ \Gamma^A_{\chi B} = \frac{r'}{r} \delta^A_B \ , \ \ \ \Gamma^A_{BC} = \Gamma^A_{BC}(S^2) \ , \label{equ:christA} \\
& & \text{($\gamma_{AB}$ is the $S^2$ metric)} \nonumber
}
we find
\Eq{ 
\label{equ:nablaV} 
\nabla_i V^j = D_i V^j + K_i^j V^0 \ , 
}
where $V$ is a four-vector, $D_i$ is the spatial covariant derivative with respect to $g_{ij}$ with constant $t$ and $K_i^j$ is the extrinsic curvature. Since the temporal component vanishes for a vector along constant-time hypersurfaces and the spatial covariant derivative with respect to $g_{ij}$ along $\delta x$ vanishes for a vector parallelly transported with respect to $g_{ij}$ along $\delta x$, this implies that
\Eq{ 
\label{equ:epsilonnablaepsilon} 
\epsilon'_q \cdot \nabla_{\delta x} \epsilon_p' = 0 
}
for $t=t_0$ and $t=t_\mathrm{ls}$. Therefore, we have found that we can set $\tilde{C}$ to be the unit matrix for our choice of the polarisation basis, and that the observed polarisation tensor can be expressed in terms of the perturbation of the polarisation tensor at the last scattering simply as
\Eq{ 
\tilde{\rho}_{pq} (t_0, \bm{x}_0, \bm{k}_0)
  = \frac{1}{2} \delta_{pq} \left(\frac{\omega}{T_\mathrm{ls}}\right) 
    + \delta \rho_{pq} (t_\mathrm{ls}, \bm{x}, \bm{k}) \ .
}
Thus, the expressions for the temperature and polarisation can be calculated as in the FLRW-universe case.

Finally, we note that we have to change the polarisation basis from the global basis to a mode-dependent one in order to express $\delta \rho_{pq} (t_\mathrm{ls}, \bm{x}, \bm{k}) $ in the above formula in terms of the mode-dependent expression for the initial condition, $\rho^{(1)}$ in \S\ref{sec:initcond}. 

From the spatial flatness at last scattering and the spherical symmetry of the background, it follows that the global polarisation basis on the last scattering surface defined above can be written as
\Eq{ 
\label{equ:epsilonprimep} 
\bm{\epsilon}'_p = \frac{1}{a} \hat{\bm{\epsilon}}_p \ , 
}
where $\hat{\bm{\epsilon}}_p$'s are vectors that have the same expression in the coordinate system in which the FLRW background metric at the last scattering is expressed as
\Eq{ 
\label{equ:ds2} 
ds^2 = -dt^2 + a(t)^2 d\bm{x} \cdot d\bm{x} \ . 
}
It is easy to find the relation between this and $\bm{\epsilon}_p(\bm{K},k)$.
It is given by the rotation matrix
\Eq{ 
\label{equ:Apq} 
A_{pq} := \bm{\epsilon}'_p (t_\mathrm{ls}, \bm{x}, k) \cdot \bm{\epsilon}_q (\bm{K}, k) = R_{pq}(\theta_\mathrm{ls}) 
}
with the angle $\theta_\mathrm{ls}(\bm{K},k)$ satisfying
\Eq{ 
\label{equ:thetals} 
\cos \theta_\mathrm{ls} = \frac{\hat{\bm{K}} \cdot \hat{\bm{z}} - (\hat{\bm{K}} \cdot \hat{\bm{k}}) ( \hat{\bm{z}} \cdot \hat{\bm{k}})}{\sqrt{1-(\hat{\bm{K}} \cdot \hat{\bm{k}})^2} \sqrt{1-(\hat{\bm{z}} \cdot \hat{\bm{k}})^2}} \ , \ \ \ 
\sin \theta_\mathrm{ls} = \frac{\hat{\bm{K}} \cdot (\hat{\bm{k}} \times \hat{\bm{z}})}{\sqrt{1-(\hat{\bm{K}} \cdot \hat{\bm{k}})^2} \sqrt{1-(\hat{\bm{z}} \cdot \hat{\bm{k}})^2}} \ .
}
The initial condition for $\delta \rho_{pq}$ at the last scattering surface is thus given by
\Eq{ 
\label{equ:deltarhopq} 
\delta \rho_{pq} (t= t_\mathrm{ls}, \bm{x}, \bm{k}) = \int d^3 \bm{K} e^{ i \bm{K} \cdot \bm{x}} A(\hat{\bm{K}} , \hat{\bm{k}}) \rho^{(1)} ( \bm{K}, \omega, \mu) A(\hat{\bm{K}}, \hat{\bm{k}})^\dag \ .
}
%

\section{Results}
\label{sec:result}

\subsection{Temperature}
\label{sec:temp}

We denote the unlensed and lensed temperature anisotropies by $\Theta$ and $\Theta'$, respectively. By the shift vector $\delta \bm{\theta}$, we define the direction $ {\bm{n}}'( {\bm{n}}) :=  {\bm{n}} + \delta \bm{\theta} $, where $ {\bm{n}}:= -  {\bm{k}}_0$ is the direction to which the observer looks. From (\ref{equ:deltathetaobs}), the components of the shift vector are $\delta \theta = D \Gamma \sin \theta$ and $\delta \varphi = 0$, where
\Eq{
\Gamma := - \int_{t_\r{ls}}^{t_0} \frac{dt}{\chi^2} 
    \left( \frac{\chi}{r}(1- \xi) + \frac{1}{S} - \frac{\chi}{r} e^{Y(t,t_0)} \right) . 
\label{equ:Gamma} 
}
Therefore, if $D=0$, then $ {\bm{n}}' =  {\bm{n}}$. With these definitions, we have
\Eq{ 
\Theta'( {\bm{n}}) = \Theta( {\bm{n}}') \ .
\label{equ:ThetaprimeTheta} 
}

The l.h.s. of \eqref{equ:ThetaprimeTheta} can be expanded as
\Eq{  
\Theta'( {\bm{n}}) = \sum_{\ell m} {\Theta'}_{\ell}^{m} Y_{\ell}^{m} ( {\bm{n}}) \ , 
\label{equ:Thetaprimeexp} 
}
while the Taylor expansion of the r.h.s. gives
\Eqr{ 
\Theta(\bm{n}') =\Theta(\bm{n}+\delta\bm{\theta})= \Theta(\bm{n})
   +\delta\bm{\theta}\cdot\nabla\Theta(\bm{n}) + \Order{D^2}.
}
Hence, $\delta\Theta^m_\ell:=\Theta'{}^m_{\ell}-\Theta^m_\ell$ can be calculated up to the first order in $D$ as
\Eqr{
\delta\Theta^m_\ell &=& D\Gamma\int d\bm{n}Y^m_\ell{}^*(\bm{n}) (\mu^2-1)\pd_\mu 
 \sum_{\ell', m'}\Theta^{m'}_{\ell'} Y^{m'}_{\ell'}(\bm{n})
 \notag\\
 &=& D\Gamma\insbra{
  (\ell - 1) \sqrt{\frac{\ell^2 - m^2}{4 \ell^2 -1 }} \Theta_{\ell - 1}^m 
   - (\ell + 2) \sqrt{\frac{(\ell + 1)^2 - m^2}{4(\ell + 1)^2 - 1}} \Theta_{\ell + 1}^m } \ ,
\label{equ:deltaThetalm2}
}
where we have used the identity\cite{Varshalovich:1988ye}
\Eq{
(\mu^2 - 1) \frac{\partial}{\partial \mu} Y_\ell^m = \ell C_{\ell + 1}^m Y_{\ell + 1}^m - (\ell + 1) C_\ell^m Y_{\ell - 1}^m \ , 
 \label{equ:shformula1} 
}
with
\Eq{ 
C_\ell^m = \sqrt{\frac{\ell^2 - m^2}{4 \ell^2 - 1}} \ . 
\label{equ:clm} 
}
%

\subsection{Polarisation}
\label{sec:pol}

As in the case of the temperature anisotropy, the lensed polarisation tensor $P'_{ab}(\bm{n})$ and the unlensed polarisation tensor $P_{ab}(\bm{n})$ are related by
\Eq{ 
P'_{ab} ( {\bm{n}}) = P_{ab} ( {\bm{n}}') \ . 
\label{equ:PprimeP} 
}
Accordingly, we obtain  the relation
\Eq{ 
{E'}_{\ell}^{m} \pm i {B'}_{\ell}^{m} 
= \sum_{\ell', m'} ({E}_{\ell'}^{m'} \pm i {B}_{\ell'}^{m'} )
 \int d{\bm{n}}\,{}_{\pm 2} Y_{\ell}^{m *} ( {\bm{n}}) (\mu^2-1)\pd_\mu \, 
 {}_{\pm 2} Y_{\ell'}^{m'} ({\bm{n}})\, 
\label{equ:PprimePlm} \ . 
}
Using the formula\cite{Varshalovich:1988ye}
\Eq{ 
(\mu^2 - 1) \frac{\partial}{\partial \mu} {}_s Y_\ell^m = \ell {}_s C_{\ell + 1}^m {}_s Y_{\ell + 1}^m - (\ell + 1) {}_s C_\ell^m {}_s Y_{\ell - 1}^m + \frac{sm}{\ell(\ell + 1)} {}_s Y_\ell^m \ , 
\label{equ:swshformula1} 
}
with
\Eq{ 
{}_s C_\ell^m = \sqrt{\frac{(\ell^2 - m^2)(\ell^2 - s^2)}{\ell^2 (4 \ell^2 - 1)}} \ , 
\label{equ:sclm} 
}
we obtain
\Eqrsub{ 
\delta{E}_\ell^m 
&=& - D \Gamma  \sqrt{\frac{((\ell+1)^2 - m^2)((\ell+1)^2 - 4)}{(\ell+1)^2 (4 (\ell+1)^2 - 1)}} (\ell + 2) E_{\ell + 1}^m 
\notag\\
&& \quad
 + D \Gamma  \sqrt{\frac{(\ell^2 - m^2)(\ell^2 - 4)}{\ell^2 (4 \ell^2 - 1)}} (\ell - 1) E_{\ell - 1}^m \nonumber \\
&& \quad
 + i D \Gamma \frac{2 m}{\ell(\ell + 1)} B_\ell^m  \ ,
\label{equ:Eprime}\\
\delta{B}_\ell^m 
&=& - D \Gamma  \sqrt{\frac{((\ell+1)^2 - m^2)((\ell+1)^2 - 4)}{(\ell+1)^2 (4 (\ell+1)^2 - 1)}} (\ell + 2) B_{\ell + 1}^m 
\notag\\
&&\quad 
 + D \Gamma  \sqrt{\frac{(\ell^2 - m^2)(\ell^2 - 4)}{\ell^2 (4 \ell^2 - 1)}} (\ell - 1) B_{\ell - 1}^m \nonumber \\
&&\quad
  - i D \Gamma \frac{2 m}{\ell(\ell + 1)} E_\ell^m \ .
\label{equ:Bprime} 
}
From these, we see that even if there exists no B-mode in the unlensed polarisation anisotropy, an off-center observer with $D\neq0$ detects non-vanishing B-modes in the lensed polarisation anisotropy.

\subsection{Correlations}
\label{sec:corr}

From these relations and the assumptions on the correlations for the unlensed initial anisotropies, \eqref{equ:s2powerspectrum} and $C^{\mathrm{TB}}_\ell =C^{\mathrm{EB}}_\ell = 0$, we immediately obtain the following formulas for the correlations among lensed anisotropies of the CMB temperature and polarisation:
\Eqrsub{ 
\langle {\Theta'}_\ell^{m *} {\Theta'}_{\ell'}^{m'} \rangle 
&=& C^{\mathrm{TT}}_\ell \delta_{\ell \ell'} \delta_{m m'} \nonumber \\
&+& D \Gamma \sqrt{\frac{(\ell + 1)^2 - m^2}{4(\ell + 1)^2 - 1}} \{ \ell C^{\mathrm{TT}}_\ell - (\ell + 2) C^{\mathrm{TT}}_{\ell + 1} \} \delta_{\ell,\ell'-1} \delta_{m m'} \nonumber \\
&+& D \Gamma \sqrt{\frac{\ell^2 - m^2}{4\ell^2 - 1}} \{ -(\ell+1) C^{\mathrm{TT}}_\ell + (\ell -1) C^{\mathrm{TT}}_{\ell - 1} \} \delta_{\ell,\ell'+1} \delta_{m m'}
 \label{equ:TT} 
\\
\langle {\Theta'}_\ell^{m *} {E'}_{\ell'}^{m' } \rangle 
&=& C^{\mathrm{TE}}_\ell \delta_{\ell \ell'} \delta_{m m'} \nonumber \\
&+& D \Gamma \sqrt{\frac{(\ell + 1)^2 - m^2}{4(\ell + 1)^2 - 1}} \left\{ \sqrt{\frac{(\ell + 1)^2 - 4}{(\ell + 1)^2}} \ell C^{\mathrm{TE}}_\ell - (\ell + 2) C^{\mathrm{TE}}_{\ell + 1} \right\} \delta_{\ell,\ell'-1} \delta_{m m'} \nonumber \\
&+& D \Gamma \sqrt{\frac{\ell^2 - m^2}{4\ell^2 - 1}} \left\{ - \sqrt{\frac{\ell^2 - 4}{\ell^2}} (\ell+1) C^{\mathrm{TE}}_\ell + (\ell -1) C^{\mathrm{TE}}_{\ell - 1} \right\} \delta_{\ell,\ell'+1} \delta_{m m'}
\notag\\
&& \label{equ:TE} 
\\
\langle {E'}_\ell^{m *} {E'}_{\ell'}^{m' } \rangle 
&=& C^{\mathrm{EE}}_\ell \delta_{\ell \ell'} \delta_{m m'} \nonumber \\
&+& D \Gamma \sqrt{\frac{((\ell + 1)^2 - m^2)((\ell + 1)^2 - 4)}{(4(\ell + 1)^2 - 1)(\ell + 1)^2}} \{ \ell C^{\mathrm{EE}}_\ell - (\ell + 2) C^{\mathrm{EE}}_{\ell + 1} \} \delta_{\ell,\ell'-1} \delta_{m m'} \nonumber \\
&+& D \Gamma \sqrt{\frac{(\ell^2 - m^2)(\ell^2 - 4)}{(4\ell^2 - 1)\ell^2}} \{ -(\ell+1) C^{\mathrm{EE}}_\ell + (\ell -1) C^{\mathrm{EE}}_{\ell - 1} \} \delta_{\ell,\ell'+1} \delta_{m m'}
\notag\\
&&
 \label{equ:EE} 
\\
\label{equ:TB} \langle {\Theta'}_\ell^{m *} {B'}_{\ell'}^{m' } \rangle 
&=& -i D \Gamma \frac{2m}{\ell(\ell + 1)} C^{\mathrm{TE}}_\ell \delta_{\ell \ell'} \delta_{m m'} 
\\
\label{equ:EB} \langle {E'}_\ell^{m *} {B'}_{\ell'}^{m' } \rangle 
&=& -i D \Gamma \frac{2m}{\ell(\ell + 1)} ( C^{\mathrm{EE}}_\ell + C^{\mathrm{BB}}_\ell ) \delta_{\ell \ell'} \delta_{m m'} 
\\
\label{equ:BB} \langle {B'}_\ell^{m *} {B'}_{\ell'}^{m' } \rangle 
&=& C^{\mathrm{BB}}_\ell \delta_{\ell \ell'} \delta_{m m'}
}

These expressions show that an off-center observer detects  non-vanishing $T-B$ and $E-B$ correlations. This is what was expected, but these correlations have  one non-trivial significant feature; they are nonzero only for the same $\ell$, while $\langle E'{}^m_\ell E'{}^{m'}_{\ell'}\rangle$ are nonzero for $\ell'=\ell\pm1$. This feature can be used as a decisive signal showing the existence of a local void in future B-mode measurement experiments with very high precision.

\subsection{Estimation}
\label{sec:estimation}

Although the main purpose of the present paper is to derive general formulas for the temperature and polarisation anisotropies in the LTB model, we give some numerical estimations for $\Gamma$ in order to get a quantitative idea on the lensing effect of a local void.  

We consider two models characterized by the curvature parameter function $\Omega_k(\chi)$ and the density parameter function $\Omega_m(\chi)$ defined by
\Eq{
\Omega_k(\chi)= -\frac{1}{H_0^2} k(\chi),\quad
\Omega_m(\chi)= \frac{2G M(\chi)}{H_0^2\chi^3},
}
where $k(\chi)$ is the curvature function appearing in  \eqref{curvaturefn:def}, and $M(\chi)$ represents the mass inside the sphere of the coordinate radius $\chi$ at present $t=t_0$.

The first model (AA model) is the void model studied by Alnes, Amarzguioui and Gr$\mbox{\o}$n in \citen{AA2006} and \citen{AAG2006}. This model is defined by
\Eqrsub{
&& \Omega_k(\chi)=\frac{\alpha}{2} \inpare{ 1-\tanh\frac{\chi-\chi_0}{\Delta \chi}},\\
&& \Omega_m(\chi)= 1-\Omega_k(\chi).
}
The second condition fixes the gauge freedom in the choice of $\chi$. This model approaches the Einstein-de Sitter model at infinity outside the void, and the ratio of the expansion rates of the universe at present at infinity and at the center, $H_\infty/H_0$ is given by $2/(3H_0 t_0)$, where $t_0$ is the age of the universe at the center. 

The second model (modified AA model) is a modification of the AA model and defined by
\Eqrsub{
&& \Omega_k(\chi)=\frac{\alpha}{2} \inpare{ 1-\tanh\frac{\chi^2-\chi_0^2}{\Delta \chi^2}},\\
&& \Omega_m(\chi)= 1-\Omega_k(\chi).
}
The integrand for $\Gamma$ in \eqref{equ:Gamma} depends on the second $\chi$-derivative of metric coefficients. Because this model is smooth at the center, the integrand is finite at the center. In contrast, the AA model has a cusp singularity in the density and curvature. Hence, the corresponding integrand has a kind of $\delta$-function type singularity. Although we can obtain a finite value for $\Gamma$ by neglecting this contribution, it comes up when we replace the model by a smoothed one. This is the reason why we considered this modified AA model.

\begin{table}[t]
\label{tbl:shiftvector}
\begin{center}
\caption{Numerical estimates of $\Gamma$ for the AA model and the modified AA model.}
\vspace{1mm}
\begin{tabular}{|c|c|c|c|}
\hline
Void radius   & Wall width &  \multicolumn{2}{|c|}{Value of $\Gamma/H_0$} \\\cline{3-4}
 $H_0\chi_0$  &  $H_0\Delta\chi$    &     AA model  & Modified AA model \\
\hline
0.0235   & 0.020 & -3.62  & 1.68 \\
0.235    & 0.20   & -0.09 & 0.30\\
0.470    & 0.40   &  0.125 & 0.303 \\
\hline
\multicolumn{4}{l}{Density contrast parameter $\alpha=0.9$}
\end{tabular}
\end{center}
\end{table}

In Table I, we give the results of numerical estimations for three model parameters for each model. Roughly speaking, $\Gamma/H_0$ is of order unity, hence $\delta\theta=\Order{H_0 D}$. However, the exact magnitude and sign are quite sensitive to the void shape and size parameters. This should be contrasted with the dipole anisotropy, which is rather insensitive to the void size and shape. More detailed analysis is under investigation.

\section{Summary and Discussion}
\label{sec:sum}
In this paper, we have developed a formulation to calculate the gravitational lensing effects on the CMB temperature and polarisation for an observer close to the center of a spherically symmetric void described by the LTB model. In particular, we have derived explicit expressions for the correlations among the anisotropies of temperature and polarisation induced by gravitational lensing in terms of an integration of known geometrical quantities along the central past light cone.

With the helps of these formulas, we have found that for an off-center observer in the local void, there appear  nonzero correlations between $T$ and $B$ and between $E$ and $B$ that are diagonal in the harmonic coefficient expression in the leading order with respect to the observer offset distance. Similar correlations arise if there exists a quintessence-type axionic field with mass in the range from $10^{-29}\r{eV}$ to $10^{-33}\r{eV}$, but in this case the magnitudes of correlations  have different dependence on $\ell$. Hence, if the correlations suggested by our result are detected in future B-mode measurement experiments with high precision, they would provide a clear signal showing the existence of a local void. 

We have also given some preliminary numerical estimations of the gravitational lensing effect. The results indicate that the B-mode amplitude produced by the lensing is around $10^{-3}$ times that of E-modes if we take into account the constraint on the off-center distance on the observer from the dipole anisotropy of the CMB temperature\cite{2010PThPh.124..163K}. This is because the effect is proportional to the distance of the observer to the symmetry center. Thus, we need next generation B-mode experiments to use this effect to test the viability of the local void model. However, the results also show that the gravitational lensing effect is very sensitive to the void profile and sizes. Hence, when it is detected, it is useful to specify a model.  Because the lensing effect becomes larger for smaller voids in general, it may be also used to detect anomalously large voids,  which are much smaller than the standard void size ($\sim$ 1Gpc) in the local void model but still statistically rare in  the $\Lambda$CDM model.

In the present paper, we have only considered the local void model, but the method developed here can be applied to more general inhomogeneous models, such as realistic universe models with voids outside us. Such extensions are under investigation.

\section{Acknowledgements}
\label{sec:acknowledge}

We would like to thank Keiki Saito and Akihiro Ishibashi for valuable discussions and help in numerical estimations.  We would also thank all participants of the workshop $\Lambda$-LTB Cosmology (LLTB2009) held at KEK from 20 to 23  October 2009 for stimulating discussions.  This work was supported by the project Shinryoiki of the SOKENDAI Hayama Center for Advanced Studies and the MEXT Grant-in-Aid for Scientific Research on Innovative Areas (No. 21111006).


%
%
%
%
%


\def\apj{Astrophys. J. }
\def\aap{Astron. Astrophys. }
\def\mnras{Mon. Not. R. Astron. Sco. }
\def\ptp{Prog. Theor. Phys.}
\def\jcap{JCAP }
\def\prd{Phys. Rev. D }
\def\jmp{J. Math. Phys.}

\bibliographystyle{jphys_n_usrt}

\end{document}